\documentclass[12pt]{article}
\usepackage{natbib,graphicx,setspace,amsmath,amssymb,subfigure,url}

\newtheorem{thm}{Theorem}
\newtheorem{lem}{Lemma}

\begin{document}

\title{Flexible Shrinkage Estimation in High-Dimensional Varying Coefficient Models}         
\author{Heng Lian\\Division of Mathematical Sciences\\School of Physical and Mathematical Sciences\\Nanyang Technological University\\Singapore 637371\\Singapore}        
\date{}          
\maketitle
\begin{abstract}
We consider the problem of simultaneous variable selection and constant coefficient identification in high-dimensional varying coefficient models based on B-spline basis expansion. Both objectives can be considered as some type of model selection problems and we show that they can be achieved by a double shrinkage strategy. We apply the adaptive group Lasso penalty in models involving a diverging number of covariates, which can be much larger than the sample size, but we assume the number of relevant variables is smaller than the sample size via model sparsity. Such so-called ultra-high dimensional settings are especially challenging in semiparametric models as we consider here and has not been dealt with before. Under suitable conditions, we show that consistency in terms of both variable selection and constant coefficient identification can be achieved, as well as the oracle property of the constant coefficients. Even in the case that the zero and constant coefficients are known a priori, our results appear to be new in that it reduces to semivarying coefficient models (a.k.a. partially linear varying coefficient models) with a diverging number of covariates. We also theoretically demonstrate the consistency of a semiparametric BIC-type criterion in this high-dimensional context, extending several previous results. The finite sample behavior of the estimator is evaluated by some Monte Carlo studies.

\textbf{keywords:} Adaptive Lasso; Extended BIC; B-spline basis; Semivarying coefficient models; Varying coefficient models;  
\end{abstract}

\section{Introduction}       
Consider a varying coefficient model \citep{hastie93}
\begin{equation}\label{eqn:model}
Y=X\beta_0(t)+\epsilon,
\end{equation}
where $X$ is a $n\times p$ covariate matrix, $\beta_0(t)=(\beta_{01}(t),\ldots,\beta_{0p}(t))^T$ is the varying coefficients and $\epsilon=(\epsilon_1,\ldots,\epsilon_n)^T$ contains the mean zero noises. For better model interpretation and efficient estimation, it is desired to identify those irrelevant covariates ($\beta_j(t)=0$) as well as covariates associated with constant coefficients ($\beta_j(t)=c$ for some constant $c$). We allow $p>>n$ but the number of nonzero coefficients is smaller than $n$ while still converging to infinity. 

For varying coefficient models, estimation can be performed based on local polynomial regression, B-spline expansion, or smoothing splines \citep{fanzhang99,fanzhang00,chiang01,huang02,huang04,eubank04}. Local polynomial regression is a most popular approach, but it requires solving many similar optimization problems on a fine grid on the support of the index variable. Thus here we choose the B-spline expansion approach.

Shrinkage estimation for variable selection has attracted much attention recently, with many contributions on the linear or parametric model \citep{tibshirani96,fan01,fan04,zou06,yuan06,yuanlin07,zou08}. Applying this approach to nonparametric or semiparametric problems is more recent, probably starting with the COSSO method \citep{lin06,zhang06} for nonparametric models. For varying coefficient models in particular,  \citep{wang09,wang08} studied the variable selection problem using kernel regression and B-spline expansion respectively, when the dimensionality is fixed. Extension to generalized semivarying coefficient models is presented in \cite{liliang08} where penalization is used for selecting predictors in the parametric component only. Studies on constant coefficient identification is comparatively scarce, and include \cite{xia04} which used cross-validation, \cite{huang02,fanhuang05} which used hypothesis testing, and \cite{leng09} which used penalization for identifying constant coefficients in the context of smoothing splines. All of these works treat fixed dimensional problems. Regularization method for variable selection with a diverging dimensionality has been investigated recently for additive models \citep{ravikumar08,meier09,huang10}. For partially linear models, \cite{xiehuang09} considered variable selection for the parametric component when dimension increases with sample size.

Based on the works mentioned above, selecting relevant variables and choosing constant coefficient in a varying coefficient model is not a new problem, but our goal here is obviously more ambitious. First, we consider a diverging number of predictors that can increase exponentially in sample size. Such a large dimension in nonparametric models has only been used in additive models as mentioned in the previous paragraph. For our semiparametric (since it will reduce to semivarying coefficient models with both nonparametric and parametric components, even when the model is correctly specified) varying coefficient models, the situation is more complicated. Second, we consider regularization method for \textit{simultaneous} variable selection and constant coefficient identification.  Given that our method can achieve both goals, a semivarying coefficient model results. The asymptotic property of semivarying coefficient models with a diverging dimensionality appears to be new and of interest in itself, even without penalization. Third, we introduce a semiparametric BIC-type criterion for automatically choosing the regularization parameters. Consistency of BIC-type criterion in the regularization framework for nonparametric models has only been shown in the case of fixed dimension \citep{wang09}. Even for linear models, consistency has been considered only in the case with $p$ increases polynomially in sample size \citep{chenchen08,wanglileng09}. All these make our theoretical investigations very challenging, due to high dimensionality and double penalty. 


Although other penalties such as SCAD can be used, here we choose the alternative adaptive group Lasso penalty. 
The advantage is that the criterion function is convex and a global optimum is guaranteed. Convexity also means the first order KKT condition is both necessary and sufficient for optimality which is the key in our proofs.  The rest of the article is organized as follows. In the next section, we present the estimation procedure using B-spline basis expansion and discuss some computational issues. Theoretical results are given in Section 3 with proofs relegated to the Appendix. Section 4 briefly discusses the choice of the initial estimator before the adaptive group Lasso penalty can be applied. Section 5 contains some simulation studies used to illustrate the performance of the estimator, and we conclude in Section 6.

\section{Penalized estimation with double adaptive Lasso penalty}
First we note that many quantities that appear in our exposition, including the dimensionality $p$, implicitly depend on $n$. Let $(X_i,Y_i,t_i),i=1,\ldots,n,$ be independent and identically distributed observations from the varying coefficient model (\ref{eqn:model}) and for simplicity we assume the index variable $t$ has a distribution supported on $[0,1]$. We use polynomial splines to approximate the coefficients. Let $\xi_0=0<\xi_1<\cdots<\xi_{K'}<1=\xi_{K'+1}$ be a partition of $[0,1]$ into subintervals $[\xi_k,\xi_{k+1}),k=0,\ldots,K'$ with $K'$ internal knots. We only restrict our attention to equally spaced knots although data-driven choice can be considered such as using the quantiles of the observed $t_i$. A polynomial spline of order $d'$ is a function whose restriction to each subinterval is a polynomial of degree $d'-1$ and globally $d'-2$ times continuously differentiable on $[0,1]$. The collection of splines with a fixed sequence of knots has a normalized B-spline basis $\{B_{1}(t),\ldots,B_{K}(t)\}$ with $K=K'+d'$. As in \cite{deboor01}, we also assume that a linear combination of basis functions $\sum_{k=1}^Ka_kB_k(t)$ is a constant $a$ if and only if $a_1=\cdots=a_K=a$, which can be achieved by making the boundary knots have multiplicity $d'$ for example. Using spline expansions, we can approximate the coefficients by $\beta_j(t)\approx \sum_k b_{jk}B_k(t)$. Note that it is possible to specify different $K$ for each coefficient but we assume they are the same for simplicity.

We are especially interested in a sparse model where many of the coefficients $\beta_{0j}$ are zeros, and in addition some coefficients are non-varying constants. To fix ideas, we assume the first $p_1$ coefficients are truly varying, the next $p_2$ coefficients are constants and the rest are zeros, and let $s=p_1+p_2\le p$ be the total number of nonzero coefficients. In order to automatically identify those special coefficients, we propose the following penalized least square estimation procedure 
\begin{equation}\label{eqn:min}
\hat{b}=\arg\min_b\frac{1}{2}\sum_i(Y_i-\sum_{j=1}^p\sum_{k=1}^KX_{ij}b_{jk}B_k(t_i))^2+n\lambda_1\sum_{j=1}^pw_{1j}||b_j||+n\lambda_2\sum_{j=1}^pw_{2j}||b_j||_c,
\end{equation}
where $\lambda_1, \lambda_2$ are regularization parameters, $w_1=(w_{11},\ldots,w_{1p})$ and $w_2=(w_{21},\ldots,w_{2p})$ are two given vectors of weights, need to be appropriately chosen in order to achieve consistency in model selection. One possible choice of these weights is obtained from an initial estimator based on group Lasso penalty (that is, equation (\ref{eqn:min}) with weights equal to 1), resulting in a globally two-step approach in estimating the coefficients. Some discussions on the initial estimator are provided in Section 4 and for now we assume the weights are already given. For the penalty terms in (\ref{eqn:min}), $||a||=(\sum_{k=1}^Ka_k^2)^{1/2}$ is the $l_2$ norm of any $K-$dimensional vector $a$ and $||a||_c=(\sum_{k=1}^K(a_k-\bar{a})^2)^{1/2}$ with $\bar{a}=\sum_{k=1}^Ka_k/K$. We note that the first penalty is used for identifying zero coefficients while the second is used for identifying constant coefficients, since $||b_j||_c=0$ if and only if $b_{j1}=\cdots=b_{jK}$. For future reference, we remark that $||a||_c$ is actually the Euclidean distance from $a$ to the linear subspace $L=\{b\mathbf{1}, b\in R\}$, where $\mathbf{1}$ is the vector with all components ones, and can thus be written equivalently as $||Q_La||$ with $Q_L$ the $K\times K$ matrix representing the projection onto the orthogonal complement of $L$.

The minimization problem can be solved by locally quadratic approximation as suggested in \cite{fan01,wang08,wang09} which is by now a rather well-known and standard algorithm. Using the notations 
\[Z_j=\left(\begin{array}{cccc}
	X_{1j}B_{1}(t_1)&X_{1j}B_{2}(t_1)&\cdots&X_{1j}B_{K}(t_1)\\
        \cdots&\cdots&\cdots&\cdots\\
	X_{nj}B_{1}(t_n)&X_{nj}B_{2}(t_n)&\cdots&X_{nj}B_{K}(t_n)\\
	\end{array}\right)_{n\times K_j},\]
$Z=(Z_1,\ldots,Z_p)$, $Y=(Y_1,\ldots,Y_n)^T$, (\ref{eqn:min}) can be written in matrix form as
\begin{equation}\label{eqn:min2}
\arg\min_b\frac{1}{2}||Y-Zb||^2+n\lambda_1\sum_{j=1}^pw_{1j}||b_j||+n\lambda_2\sum_{j=1}^pw_{2j}||b_j||_c.
\end{equation}
The locally quadratic approximation approach iteratively solves
\[
\arg\min_b\frac{1}{2}||Y-Zb||^2+n\lambda_1\sum_{j=1}^pw_{1j}||b_j||^2/||b_j^{(0)}||+n\lambda_2\sum_{j=1}^pw_{2j}||b_j||_c^2/||b_j^{(0)}||_c,
\]
with $b^{(0)}$ the current estimate. However, with double penalties, we need to keep track of both zero coefficients as well as constant coefficients during the iterative process, making the implementation slightly more complicated than usual. The details are omitted here. 

In practice, we need to choose some parameters including the spline order $d'$, the number and positions of the knots of the spline basis as well as the two regularization parameters. To ease the computational burden, we fix $d'=4$ and $K=10$ with equally spaced knot sequence in our implementation and choose only $\lambda_1$ and $\lambda_2$ based on data. This strategy is well known in the functional smoothing/functional data analysis literature, where the number of knots is chosen to be sufficiently large to reduce bias in function approximation since the variance can be effectively controlled by subsequent penalization (see for example Chapter 5 of \cite{ramsay05} for a detailed illustration of this effect in the functional smoothing context). It is also possible to position the knots based on sample quantiles of the observed index variable, but since choosing optimal knots is not the focus of the paper we will only use equally spaced knots for simplicity.

We use a BIC-type criterion to select simultaneously $\lambda_1$ and $\lambda_2$, given by
\begin{equation}\label{eqn:bic}
BIC_{\lambda}=\log\{\frac{1}{n}||Y-Z\hat{b}_\lambda||^2\}+d_1\frac{\log n}{n}C_n+d_2\frac{\log(n/K)}{n/K}C_n,
\end{equation}
where $\hat{b}_\lambda$ is the minimizer of (\ref{eqn:min2}) given $\lambda=(\lambda_1,\lambda_2)$, $d_1$ is the number of coefficients estimated as nonzero constants and $d_2$ is the number of coefficients estimated as truly varying. We will show later that the BIC is consistent in model selection if $C_n=\Omega(\sqrt{\log (pK)})$ and $C_n\log(n/K)/(n/K)\rightarrow 0$ under some additional assumptions, where the notation $a_n=\Omega(b_n)$ means $b_n=O(a_n)$. We will use $C_n=\sqrt{\log (pK)}$ in our simulations which produces reasonable results. Although it is unsatisfactory that $C_n$ must be chosen in a somewhat arbitrary way, the same problem appeared in \cite{wanglileng09} in which some arbitrary value (among many possibilities) that satisfies their theoretical conditions is picked and its performance is verified using Monte Carlo examples. We will refer to the criterion (\ref{eqn:bic}) with $C_n=\sqrt{\log (pK)}$ as the extended BIC (EBIC) following \cite{chenchen08}, while with $C_n=1$ we obtain the ordinary BIC.

\section{Asymptotic results}
We first introduce the following notations. Let $Z^{(1)}$ be the $n\times p_1K$ submatrix of $Z$ containing the columns corresponding to truly varying coefficients, and similarly let $Z^{(2)}$ be the submatrix corresponding to constant coefficients and $Z^{(3)}$ the submatrix corresponding to zero coefficients. In the same spirit, we can define $X^{(1)}, X^{(2)}, X^{(3)}$ as suitable submatrices of $X$, with the corresponding random variables denoted by $x^{(1)}, x^{(2)}, x^{(3)}$. Similar notations are also applied to vectors $b$ and $\beta(t)$. 

Let $\mathcal{G}$ denote the subspace of functions on $R^{p_1}\times [0,1]$
\begin{eqnarray*}
  \mathcal{G}&:=&\{g(x^{(1)},t): g(x^{(1)},t)={x^{(1)T}}h(t), h(t)=(h_1(t),\ldots,h_{p_1}(t))^T \\
	&&\mbox{ with some functions } h_j(t) \mbox{ and } E\sum_{j=1}^{p_1}{x^{(1)2}_{j}}h_j^2(x_1,t)<\infty\},
\end{eqnarray*}
and for any random variable $w$ with $E(w^2)<\infty$, let $E_\mathcal{G}(w)$ denote the projection of $w$ onto $\mathcal{G}$ in the sense that 
\[E\{(w-E_\mathcal{G}(w))(w-E_\mathcal{G}(w))\}=\inf_{g\in \mathcal{G}}E\{(w-g(x^{(1)},t))(w-g(x^{(1)},t))\}.
\]
Definition of $E_\mathcal{G}(w)$ trivially extends to the case $w$ is a random vector by componentwise projection. 

In the theoretical studies of our estimator, we will use the decomposition 
\begin{equation}\label{eqn:decomp}
x^{(2)}=\theta(x^{(1)},t)+u=\theta(x^{(1)},t)-g(x^{(1)},t)+g(x^{(1)},t)+u,
\end{equation}
with $\theta(x^{(1)},t)=E(x^{(2)}|x^{(1)},t)$, $g(x^{(1)},t)=E_\mathcal{G}(x^{(2)})$. Note that since the conditional expectation $E(x^{(2)}|x^{(1)},t)$ can be interpreted as projection onto the space $\{h(x^{(1)},t), Eh^2<\infty \}$ of which $\mathcal{G}$ is a subspace, we see that we also have $g(x^{(1)},t)=E_\mathcal{G}(\theta(x^{(1)},t))$. Let $\Xi=E\{(x^{(2)}-g(x^{(1)},t))(x^{(2)}-g(x^{(1)},t))^T\}$ which can be considered as the residual variance of $x^{(2)}$ after projection.

For adaptive group Lasso penalty in (\ref{eqn:min2}), the weights $w_{1j}, s+1\le j\le p$ are associated with the zero coefficients and $w_{2j}, p_1+1\le j\le p$ are associated with constant (including zero) coefficients. Asymptotically, these weights do not appear in the convergence rates if we can consistently select the true model. Thus it makes sense for our asymptotic investigation to define $||w'_1||=(\sum_{j=1}^sw_{1j}^2)^{1/2}$ and $||w'_2||=(\sum_{j=1}^{p_1}w_{2j}^2)^{1/2}$ which will appear in the convergence rates.

First we consider the case where covariates corresponding to zero and constant coefficients are known to us. In this case, we have a ``regularized oracle estimator" $(\hat{b}^{(1)},\hat{\beta}^{(2)})$ obtained from minimizing the following functional
\begin{eqnarray}
Q(b^{(1)},\beta^{(2)})&=&\frac{1}{2}||Y-Z^{(1)}b^{(1)}-X^{(2)}\beta^{(2)}||^2+n\lambda_1\sum_{j=1}^{p_1}w_{1j}||b_j^{(1)}||\nonumber\\
&&+n\lambda_2\sum_{j=1}^{p_1}w_{2j}||b_j^{(1)}||_c+n\lambda_1\sqrt{K}\sum_{j=p_1+1}^{s}w_{1j}|\beta_j^{(2)}|,\label{eqn:roe}
\end{eqnarray}
where $b^{(1)}$ is a $p_1K$ dimensional vector corresponding to the truly varying coefficients and $\beta^{(2)}=(\beta^{(2)}_{p_1+1},\ldots,\beta^{(2)}_s)^T$ are the constant coefficients. The extra $\sqrt{K}$ in the penalty above is due to that $||b_j||=\sqrt{K}|\beta_j|$ when $b_{j1}=\ldots=b_{jK}=\beta_j$. 

We will consider rates of convergence as well as asymptotic normality of the resulting estimator. Note that our results for the minimizer of (\ref{eqn:roe}) cover the unpenalized case $\lambda_1=\lambda_2=0$ and thus provide some asymptotic analysis of semivarying coefficient models with diverging dimensionality, which is of independent interests\

The conditions required for our theoretical results on the regularized oracle estimator are listed here.
\begin{itemize}
\item[(c1)] The covariates have finite fourth moments, $\max_{j}EX_{ij}^4<\infty$, and the eigenvalues of $E\{(x^{(1)T},x^{(2)T})^T(x^{(1)T},x^{(2)T})\}$ are bounded away from zero and infinity. 
\item[(c2)] The noises $\epsilon_{i}$ are independent of covariates, have mean zero, variance $\sigma^2$, and finite fourth moment.
\item[(c3)] The index variable $t$ has a density bounded away from $0$ and infinity on $[0,1]$.
\item[(c4)] For $1\le j\le p_1$, $\beta_{0j}(t)$ satisfies a Lipschitz condition of order $d>1/2$: $|\beta_{0j}^{(\lfloor d\rfloor)}(t)-\beta_{0j}^{(\lfloor d\rfloor)}(s)|\le C|s-t|^{d-\lfloor d\rfloor}$, where $\lfloor d \rfloor$ is the biggest integer strictly smaller than $d$ and $\beta_{0j}^{(\lfloor d\rfloor)}(t)$ is the $\lfloor d\rfloor$-th derivative of $\beta_{0j}(t)$. The order of the B-spline used satisfies $d'\ge d+2$.
\item[(c5)] $Ks/n\rightarrow 0, s/K^{2d}\rightarrow 0, (\lambda_1^2||w'_1||^2+\lambda_2^2||w'_2||^2)K\rightarrow 0$.
\item[(c6)] The eigenvalues of $\Xi$ are bounded away from zero and infinity.
\item[(c7)] In the decomposition (\ref{eqn:decomp}), each component of $g(x^{(1)},t)$ can be written in the form $\sum_{j=1}^{p_1}x^{(1)}_{j}h_j(t)$ for some $h_j$. We assume all $h_j$ satisfy a Lipschitz condition of order $d_g>1/2$: $|h_j^{(\lfloor d_g\rfloor)}(t)-h_j^{(\lfloor d_g\rfloor)}(s)|\le C|s-t|^{d_g-\lfloor d_g\rfloor}$. The order of the B-spline used satisfies $d'\ge d_g+2$.
\item[(c8)] $Ks^2/n\rightarrow 0, s^2/K^{2d_g}\rightarrow 0,$ and $\sqrt{ns}K^{-(d+d_g)}\rightarrow 0$.
\end{itemize}

In condition (c1), we only require the eigenvalues of the second moment matrix of covariates associated with nonzero coefficients are bounded away from zero and infinity. Conditions (c2)-(c4) are standard. The convergence rate (\ref{eqn:ratebeta}) below would be void without condition (c5). Other conditions are used in showing the faster convergence rate of the parametric component in (\ref{eqn:roe}), which is the more difficult part of the proof. (c6) and (c7) imply that $x^{(2)}$ is not in $\mathcal{G}$ and its projection onto $\mathcal{G}$ is smooth enough. These conditions are similar to Assumption (A2) and Condition 1 in \cite{xiehuang09} respectively for high-dimensional partially linear models. From the rates obtained below, if $\lambda_1=\lambda_2=0$ (or small enough), the optimal number of knots in spline expansion is $K\sim n^{1/(2d+1)}$ as usual.

\begin{thm}\label{thm:rates} (Convergence rates) Under conditions (c1)-(c5), the nonparametric component of the minimizer of (\ref{eqn:roe}), $\hat{b}^{(1)}$, satisfies 
\[||\hat{b}^{(1)}-b^0||^2=O\left(\frac{K^2s}{n}+\frac{s}{K^{2d-1}}+(\lambda_1^2||w'_1||^2+\lambda_2^2||w'_2||^2)K^2\right),\]
where $b^0$ is any vector satisfying $||\beta_{0j}(t)-\sum_k b^0_{jk}B_k(t)||=O(K^{-2d})$. As an immediate corollary, 
\begin{equation}\label{eqn:ratebeta}
\sum_{j=1}^{p_1}||\hat{\beta}^{(1)}_j(t)-\beta_{0j}(t)||^2=O\left(\frac{Ks}{n}+\frac{s}{K^{2d}}+(\lambda_1^2||w'_1||^2+\lambda_2^2||w'_2||^2)K\right),
\end{equation}
where $\beta_{0j}(t)$ denotes the true coefficients and $\hat{\beta}_j^{(1)}(t)=\sum_k\hat{b}^{(1)}_{jk}B_k(t)$.

For the parametric part, under additional assumptions (c6)-(c8), we have the faster rate
\[
\sum_{j=p_1+1}^s|\hat{\beta}_j^{(2)}-\beta_{0j}|^2=O\left(\frac{s}{n}+(\lambda_1^2||w'_1||^2+\lambda_2^2||w'_2||^2)K\right).
\]

\end{thm}

The following conditions are assumed for asymptotic normality of the parametric component.
\begin{itemize}
\item[(c9)] $s/K^d\rightarrow 0, \sqrt{ns^3}K^{-(d+d_g)}\rightarrow 0.$
\item[(c10)] $\sqrt{nKs}(\lambda_1||w'_1||+\lambda_2||w'_2||)\rightarrow 0$.
\end{itemize}
\begin{thm}\label{thm:an} (Asymptotic normality) Let $A_n$ be a deterministic $m\times p_2$ matrix with $m$ an integer that does not change with $n$, and $\Sigma_n=A_n\Xi^{-1}A_n^T$ ($\Xi$ is defined below (\ref{eqn:decomp})). Under conditions (c1)-(c10),
\[\sqrt{n}\Sigma_n^{-1/2}A_n(\hat{\beta}^{(2)}-\beta_0^{(2)})\rightarrow N(0,\sigma^2 I_m) \mbox{ in distribution },\]
where $I_m$ is the $m\times m$ identity matrix.
\end{thm}

We will now show that the estimator from (\ref{eqn:min2}) is exactly equal to the regularized oracle estimator from (\ref{eqn:roe}) with probability converging to 1. In particular, this immediately gives the same convergence rates as well as asymptotic normality as in Theorems \ref{thm:rates} and \ref{thm:an} for the estimator even when the position of the zero and constant coefficients are unknown. In order for the adaptive group Lasso estimator to identify the correct model, we need to make sure the weights $w_{1j}, s+1\le j\le p$ associated with zero coefficients and weights $w_{2j}, p_1+1\le j\le p$ associated with constant coefficients are big enough to force sufficient penalty. The following two conditions make this requirement exact. Our conditions are stated for direct use in the proof of the theorem and seem complicated. We will make the conditions more explicit in Section 4 and show that these conditions can be naturally satisfied.
\begin{itemize}
\item[(c11)] $\sqrt{n/K}\{\sqrt{\log(pK)}+\sqrt{Ks+ns/K^{2d}}+\sqrt{nK}(\lambda_1||w'_1||+\lambda_2||w'_2||)\}=o(n\lambda_2w_{2j}), p_1+1\le j\le p.$
\item[(c12)] $\sqrt{n/K}\{\sqrt{\log(pK)}+\sqrt{Ks+ns/K^{2d}}+\sqrt{nK}(\lambda_1||w'_1||+\lambda_2||w'_2||)\}=o(n\lambda_1w_{1j}), s+1\le j\le p.$
\end{itemize} 
\begin{thm}\label{thm:oracle} Assume conditions (c11) and (c12) as well as those in Theorem \ref{thm:rates}. Suppose $(\hat{b}^{(1)},\hat{\beta}^{(2)})$ solves the problem (\ref{eqn:roe}). Define $\hat{b}=(\hat{b}^{(1)},\hat{b}^{(2)},\hat{b}^{(3)})$ with $\hat{b}^{(2)}_{jk}=\hat{\beta}^{(2)}_j, p_1+1\le j\le s, 1\le k\le K$ and $\hat{b}^{(3)}_{jk}=0, s+1\le j\le p, 1\le k\le K$. Then with probability approaching 1, $\hat{b}$ is the solution of the original problem $(\ref{eqn:min2})$. As a corollary, the rates of convergence of $\hat{b}$ is the same as those stated in Theorem \ref{thm:rates} and asymptotic normality of the estimated constant coefficients holds under the additional conditions assumed in Theorem \ref{thm:an}.
\end{thm}

Finally, we consider the consistency of the BIC-type criterion. Since we consider ultra-high dimensional problems here with $p>>n$, for technical reasons, we will assume that the number of nonzero coefficients $s=O(1)$ does not increase with $n$, and that we only select among potential models with dimension upper bounded by a known integer $S$. Although restrictive in some situations, this assumption is satisfied, say, when we know that only a small number of predictors are relevant even as we collect more predictors as sample size increases, and we have an a priori bound on the number of relevant covariates. In the case of parametric models, even with $p$ only increasing polynomially in $n$, \cite{chenchen08} also makes this assumption. We need the following conditions. 
\begin{itemize}
\item[(c13)] Both $\inf_{1\le j\le p_1}||\beta_{0j}(t)||_c$ and $\inf_{p_1+1\le j\le s}|\beta_{0j}|$ are bounded away from zero.
\item[(c14)] $K\sim n^{1/(2d+1)}, C_n=\Omega(\sqrt{\log(pK)}), C_n\log(n/K)/(n/K)\rightarrow 0$.
\end{itemize}

\begin{thm}\label{thm:bic} If the number of nonzero coefficients $s$ does not increase with $n$, and we only consider models with at most $S$ (also does not increase with $n$) nonzero coefficients with $s\le S$. Under conditions (c13) and (c14), in addition to those assumed in Theorem \ref{thm:rates} and Theorem \ref{thm:oracle}, the BIC-type criterion (\ref{eqn:bic}) will correctly identify the nonzero coefficients and the constant coefficients with probability approaching 1.
\end{thm}

\section{Initial estimator with Lasso penalty} 
In the adaptive Lasso penalty, conditions (c11) and (c12) require that the weight $w_{1j}$ is large  for zero coefficient and small for nonzero ones, and similar requirements for $w_{2j}$ are imposed. Following \cite{zou06} where the adaptive Lasso is first proposed, we set $w_{1j}=1/||\tilde{b}_j||$ and $w_{2j}=1/||\tilde{b}_j||_c$ using an initial estimator $\tilde{b}$ obtained by minimizing the least square with group Lasso penalty
\[
\tilde{b}=\arg\min_b\frac{1}{2}||Y-Zb||^2+n\lambda_0\sum_{j=1}^p ||b_j||.
\]
Note that to obtain the initial estimator, it is only necessary to use a single penalty term.

\begin{thm}\label{thm:init} Under conditions (c1)-(c5), if $\lambda_0\ge C\sqrt{s\log(pK)/n}$ for sufficiently large $C>0$, all coefficients except $Ms$ of them are estimated as zeros where $M$ is a finite constant $M>1$. In addition, we have the convergence rate
\[||\tilde{b}-b^0||^2=O\left(\frac{K^2s^2\log(pK)}{n}+\frac{s}{K^{2d-1}}+\lambda_0^2K^2s\right),\]
where $b^0$ contains the coefficients in the optimal approximation of $\beta_{0j}, 1\le j\le s$ in spline basis expansion.
\end{thm}

Compared with Theorem \ref{thm:rates}, the extra factor $s\log(pK)$ in the convergence rate is due to that we do not have a priori knowledge on the nonzero components as in Theorem \ref{thm:rates}, and the logarithmic factor turns out to be the resulting cost (also see the proof of Theorem \ref{thm:oracle} where similar logarithmic factors appear in conditions (c11) and (c12)).

Equipped with the initial estimator which gives us the weights in (\ref{eqn:min2}), we will demonstrate that various conditions imposed in the previous section can be satisfied. First we fix  $\lambda_0=C\sqrt{s\log(pK)/n}$ and $K\sim n^{1/(2d+1)}$. Then the convergence rate of $||\tilde{b}-b||$ in Theorem \ref{thm:init} for the group Lasso estimator is $\sqrt{K^2s^2\log(pK)/n}=o(\sqrt{K})$ if we assume $Ks^2\log(pK)/n\rightarrow 0$, which is stronger than (c8). Suppose that condition (c12) on the true coefficients is satisfied, then the weights satisfy $w_{1j}=O(1/\sqrt{K}), 1\le j\le s$ and $w_{1j}=\Omega(\sqrt{n/(K^2s^2\log(pK))}), s+1\le j\le p$. Similarly $w_{2j}=O(1/\sqrt{K}), 1\le j\le p_1$ and $w_{2j}=\Omega(\sqrt{n/(K^2s^2\log(pK))}), p_1+1\le j\le p$.

If $\lambda_1,\lambda_2=O(\sqrt{K/n})$, then $(\lambda_1^2||w'_1||^2+\lambda_2||w'_2||^2)K^2=O(K^2s/n)$ and thus the last term in the convergence rate of $||\hat{b}-b||^2$ in Theorem \ref{thm:rates} can be ignored. If furthermore 
\begin{equation}\label{eqn:lambdasmall}
\lambda_1,\lambda_2=o(\frac{1}{\sqrt{n}s}),
\end{equation}
then condition (c10) is satisfied.

To fix ideas, suppose now $\log p=n^q$ with $0<q<1$. Conditions (c11) and (c12) impose that 
\begin{equation}\label{eqn:lambdabig}
\lambda_1,\lambda_2>>\max\{\frac{K^{1/2}sn^q}{n}, \frac{Ks^{3/2}n^{q/2}}{n}\}.
\end{equation}
If $s=O(1)$ (although not necessary), there exists $\lambda_1,\lambda_2$ that satisfies both (\ref{eqn:lambdasmall}) and (\ref{eqn:lambdabig}) if $q<d/(2d+1)$. 

To make the initial estimator effectively usable as weights, the regularization parameter $\lambda_0$ must be large enough so that many zero coefficients are correctly identified, but small enough that it still obtains reasonable convergence rates. We do not have corresponding theoretical results on how to choose $\lambda_0$ based on data. In our simulations, we use both ordinary BIC and EBIC to select this smoothing parameter. It is found that while EBIC is better at identifying the correct model when using the group Lasso penalty, BIC is more desirable in this initial step when considering our adaptive group Lasso penalty.

\section{Simulation}
In this section we use some simulations to evaluate the finite sample performance of the adaptive group Lasso in variable selection and constant coefficient identification. The datasets are generated from model (\ref{eqn:model}) with sample size $n=100$ and noises $\epsilon_i\sim N(0,0.1)$. The index variable $t$ is sampled uniformly on $[0,1]$, and the predictors are $X_{i1}=1$ with other $X_{ij}$'s marginally standard normal with within subject correlations $Cov(X_{ij_1},X_{ij_2})=(1/2)^{|j_1-j_2|}$. The first three coefficient functions are truly varying with
\begin{eqnarray*}
\beta_1(t)&=&3\sin (2\pi t),\\
\beta_2(t)&=&8t(1-t),\\
\beta_3(t)&=&\cos[ (2\pi t)^2].\\
\end{eqnarray*}
There are 6 constant coefficients specified as $\beta_4=\beta_5=1.5$, $\beta_6=\beta_7=0.5$ and $\beta_8=\beta_9=0.1$. All other coefficients are set to be zero. Since we focus on high-dimensional models here, we consider both $p=50$ and $p=150$. For both scenarios, $500$ datasets are generated and fitted. We compare adaptive group Lasso with group Lasso and also compare the effects of using ordinary BIC with extended BIC. We fix the number of spline basis $K$ to be $10$ which is sufficiently flexible to approximate the varying coefficients. For group Lasso estimator, we use both BIC and EBIC for model identification. We also consider adaptive group Lasso estimator when group Lasso estimator (using ordinary BIC) is used as the initial estimator, with $\lambda_1$ and $\lambda_2$ chosen by either ordinary BIC or EBIC.  In Table \ref{tab:model}, we show the number of identified zero and constant coefficients by different methods, with information criterion used in each case indicated in brackets. For example, the row indicated as aglasso(BIC-EBIC) shows the results for the adaptive group Lasso estimator when BIC is used in choosing $\lambda_0$ for the initial group Lasso estimator and EBIC is used in choosing smoothing parameters for the final estimator. We see that when EBIC is used for the initial estimator, some nonzero coefficients are incorrectly identified as zeros. Note that these mistakes cannot be corrected by the subsequent adaptive group Lasso estimator. On the other hand, if BIC is used for the initial estimator, although many zero coefficients are identified as varying, these mistakes can however be corrected by the final estimator. This is actually why we don't consider the combinations EBIC-BIC and EBIC-EBIC for the final estimator in our simulations. Another important conclusion to be drawn from the table is that model selection using BIC-EBIC is better than using BIC-BIC. For example, when $p=50$, the number of zero coefficients is $41$ and on average $40.26$ of them are identified using BIC-EBIC while only $36.84$ of them are identified using BIC-BIC (i.e., more false positives). BIC-EBIC also works better for identifying the constant coefficients.

In Table \ref{tab:error}, we present the estimation errors (in $L_2$ norm) for some of the coefficients. Note that based on the true model, $\beta_1,\beta_2,\beta_3$ are varying coefficients, $\beta_4,\beta_6,\beta_8$ are constants and $\beta_{10}$ is actually zero. We also show in the last column of the table the estimation error of the oracle estimator where the true model is known and no penalization is used, with $K$ selected by GCV criterion (note that here we need to choose $K$ based on data since there is no subsequent penalization that reduces the variance of the estimator if $K$ is fixed to be sufficiently large). From the table, we see that adaptive group Lasso estimator in general performs better than group Lasso estimator and for adaptive group Lasso estimator, using BIC-BIC and BIC-EBIC produces similar results (note that this is in terms of estimation error only, and BIC-EBIC is better for identifying the true model).

\begin{table}
\caption{Model selection results of different estimators based on 500 replications, with $n=100$.\label{tab:model}}
\vspace{0.1in}
\centering
{
\begin{tabular}{cccccc}
\hline\hline
&&\multicolumn{2}{c}{Avg \# of zero coefficients}&\multicolumn{2}{c}{Avg \# of const. coefficients}\\
\cline{3-4} \cline{5-6}
&&correct&incorrect&correct&incorrect\\
\hline
$p=50$&glasso(BIC)&5.94&0&0&0\\
&glasso(EBIC)&40.78&4.72&0&0\\
&aglasso(BIC-BIC)&36.84&0.01&3.98&3.12\\
&aglasso(BIC-EBIC)&40.26&0.02&5.54&0.74\\
$p=100$&glasso(BIC)&68.17&0.07&0&0\\
&glasso(EBIC)&127.36&2.1&0&0\\
&aglasso(BIC-BIC)&133.17&0.07&3.5&6.8\\
&aglasso(BIC-EBIC)&139.9&0.37&4.43&1.13\\
\hline
\end{tabular}}
\end{table}

\begin{table}
\caption{Estimation errors of different estimators based on 500 replications, with $n=100$.\label{tab:error}}
\vspace{0.1in}
\centering
{
\begin{tabular}{ccccccc}
\hline\hline
&&\multicolumn{2}{c}{glasso}&\multicolumn{2}{c}{aglasso}&oracle\\
\cline{3-4} \cline{5-6}
&&BIC&EBIC&BIC-BIC&BIC-EBIC&\\
\hline
$p=50$& $\beta_1$ & 0.0860& 1.2534 & 0.0441 & 0.0445 & 0.0399  \\
     & $\beta_2$ & 0.1076& 1.3801  & 0.0542 & 0.0671 & 0.0465  \\
     & $\beta_3$ & 0.1461& 0.5752  & 0.0671 & 0.0773 & 0.0491  \\
     & $\beta_4$ & 0.1078& 1.3779  & 0.0361 & 0.0197 & 0.0148  \\
     & $\beta_6$ & 0.0792& 0.4718  & 0.0295 & 0.0196 & 0.0171  \\
     & $\beta_8$ & 0.0460& 0.0998  & 0.0364 & 0.0242 & 0.0153  \\
     & $\beta_{10}$ & 0.0188& 0.0003 & 0.0060 & 0.0023 & 0.0000  \\
\\
$p=150$& $\beta_1$ & 0.1568&  0.5449 & 0.0571 & 0.0635 &   \\
     & $\beta_2$ & 0.1541&  0.5415 & 0.0894 & 0.0926 &   \\
     & $\beta_3$ & 0.2452&  0.3879 & 0.1221 & 0.1401 &  \\
     & $\beta_4$ & 0.1540&  0.5295 & 0.0439 & 0.0364 &  \\
     & $\beta_6$ & 0.1001&  0.2164 & 0.0387 & 0.0326 &  \\
     & $\beta_8$ & 0.0557&  0.0814 & 0.0493 & 0.0492 &   \\
     & $\beta_{10}$ & 0.0129&  0.0058 & 0.0032 & 0.0022 &   \\

\hline
\end{tabular}}
\end{table}

\section{Conclusion} 
In this paper we proposed an estimation method for identifying zero coefficients and constant coefficients simultaneously for high-dimensional varying coefficient models. The high dimensionality and the double penalties used to achieve both goals make the theoretical analysis harder than previously proposed models. We demonstrated convergence rates and asymptotic normality of the constant coefficients, and proposed semiparametric BIC as a consistent model selection tool.

One possible extension of the current work is to consider generalized varying coefficient models. Variable selection for such models has been considered in \cite{liliang08} based on local linear regression for fixed dimension. However, in their procedure, undersmoothing of the varying coefficients is necessary for efficient estimation of the parametric component. It is expected that such undersmoothing is not necessary for spline based method that estimates both components simultaneously. 

\section*{Appendix}
In some of the proofs below we will make use of some simple properties of the subdifferential and thus we first mention these properties here. For a vector $b$, the subdifferential of its $l_2$ norm is 
\[
\partial ||b||=\left\{\begin{array}{cc}
			b/||b|| & \mbox { if } b\neq 0\\
			\mbox{ some } a \mbox{ with } ||a||\le 1& \mbox{ if } b=0.
			\end{array}\right.
\]
Note that when $b=0$ the subdifferential is not unique but we still use $\partial||b||$ to denote some subdifferential since its specific value has no sigficance in our proofs. Slightly more generally, for any matrix $A$,
\[
\partial ||Ab||=\left\{\begin{array}{cc}
			A^TAb/||Ab|| & \mbox { if } Ab\neq 0\\
			A^Ts \mbox{  for some } a \mbox{ with } ||a||\le 1& \mbox{ if } Ab=0.
			\end{array}\right.
\]

\textit{Proof of Theorem \ref{thm:rates}.} The convergence rate for the nonparametric component is relatively easy to show. Instead of showing the rates for the regularized oracle estimator, we consider instead the minimizer $\hat{b}$ of the following functional
\[Q'(b)=\frac{1}{2}||Y-Zb||^2+n\lambda_1\sum_{j=1}^s w_{1j}||b_j||+n\lambda_2\sum_{j=1}^{p_2}w_{2j}||b_j||_c,\]
 where only for the proof of Theorem \ref{thm:rates} we set $Z=(Z^{(1)},Z^{(2)})$. That is, one knows the zero coefficients but does not constrain the truly constant coefficients to be constants. This makes the notation simpler. The convergence of regularized oracle estimator follows exactly the same lines.

Suppose $\beta_{nj}(t)=\sum_{k=1}^{K} b^0_{jk}B_{k}(t)$ is the best approximating spline for $\beta_{0j}(t)$ with $||\beta_{nj}-\beta_{0j}||^2=O(K^{-2d})$. By the definition of $\hat{b}$, we have
\begin{eqnarray}\label{eqn:2}
0&\ge&Q'(\hat{b})-Q'({b^0})\nonumber\\
&\ge&||Y-Z\hat{b}||^2/2-||Y-Zb^0||^2/2-n\lambda_1\sum_{j=1}^sw_{1j}||\hat{b}_j-b^0_j||-n\lambda_2\sum_{j=1}^{p_1}w_{2j}||\hat{b}_j-b^0_j||\nonumber\\
&=&(Y-Zb^0)^TZ(b^0-\hat{b})+||Z(b^0-\hat{b})||^2/2-n\lambda_1\sum_{j=1}^sw_{1j}||\hat{b}_j-b^0_j||-n\lambda_2\sum_{j=1}^{p_1}w_{2j}||\hat{b}_j-b^0_j||,\nonumber
\end{eqnarray}
where in the second inequality above we used the property $||a||_c\le||a||$ for any vector $a$. 

Let $\eta=P_Z(Y-Zb^0)$, where $P_Z=Z(Z^TZ)^{-1}Z^T$, be the projection of $Y-Zb^0$ onto the columns of $Z$, then Lemma \ref{lem:eta} shows that $||\eta||^2=O_p(Ks+ns/K^{2d})$. Using the Cauchy-Schwartz inequality, the above displayed equation can be continued as 
\begin{equation}\label{eqn:ratebound}
0\ge-|O_p(Ks+ns/K^{2d})|+\frac{1}{4}||Z(b^0-\hat{b})||^2-n\lambda_1\sum_{j=1}^sw_{1j}||\hat{b}_j-b^0_j||-n\lambda_2\sum_{j=1}^{p_1}w_{2j}||\hat{b}_j-b^0_j||.\\
\end{equation}
Using now Lemma A.1 in \cite{wang08} together with condition (c1), which implies that $||Z(b^0-\hat{b})||^2\sim (n/K)||b^0-\hat{b}||^2$, and using the Cauchy-Schwartz inequality $n\sum_j\lambda_1w_{1j}||\hat{b}_j-b^0_j||\le (CKn/4)\sum_j(\lambda_1w_{1j})^2+(n/CK)||b^0-\hat{b}||^2$ with a sufficiently large $C>0$ (similarly for $n\sum_j\lambda_2w_{2j}||\hat{b}_j-b^0_j||$), (\ref{eqn:ratebound}) implies $||\hat{b}-b^0||^2=O_p(K^2s/n+s/K^{2d-1}+(\lambda_1^2\sum_{j=1}^sw_{1j}^2+\lambda_2^2\sum_{j=1}^{p_1}w^2_{2j})K^2)$. The convergence rate for $\sum_{j=1}^{p_1}||\hat{\beta}^{(1)}_j-\beta_{0j}||^2$ is obtained from the well-know relation $||\sum_ka_kB_k(t)||^2\sim ||a||^2/K$ for any $a=(a_1,\ldots,a_K)$.

Now consider the faster convergence rate of the parametric components in the regularized oracle estimator, which we show by profiling out $b^{(1)}$ in (\ref{eqn:roe}).
For any given $\beta$, let $\hat{b}(\beta)$ be the minimizer of (\ref{eqn:roe}) when $\beta$ is fixed. Again, for ease of notation, we write $b^{(1)}$ simply as $b$, $\beta^{(2)}$ as $\beta$, $Z^{(1)}$ as $Z$, and $X^{(2)}$ as $X$. By the KKT condition, we know that $\hat{b}(\beta)$ satisfies 
\[-Z_j^T(Y-Z{b}-X\beta)+n\lambda_1w_{1j}\partial||{b}_j||+n\lambda_2w_{2j}\partial||b_j||_c, j=1,\ldots,p_1.\]
From the above expression we get 
\begin{equation}\label{eqn:defv}
\hat{b}(\beta)=(Z^TZ)^{-1}Z^T(Y-X\beta)+(Z^TZ)^{-1}v(\beta),
\end{equation}
where $v(\beta)$ is a $p_1$-dimensional vector with its $j$-th component given by $n\lambda_1w_{1j}\partial||\hat{b}_j(\beta)||+n\lambda_2w_{2j}\partial||\hat{b}_j(\beta)||_c$.

Let $\beta_0$ be the true parameter and set $\hat{\beta}=\beta_0+\gamma_1u$ with $\gamma_1=C(\sqrt{s/n}+\sqrt{K}(\lambda_1||w'_1||+\lambda_2||w'_2||))$ for some $C>0$, and $||u||=1$. We will show that $\inf_{||u||=1}Q(\hat{b}(\hat{\beta}),\hat{\beta})-Q(\hat{b}(\beta_0),\beta_0)>0$ with probability approaching 1 for $C$ large enough and the result will follow.

Using the closed form expression for $\hat{b}(\beta)$, we get
\begin{eqnarray}
&&Q(\hat{b}(\hat{\beta}),\hat{\beta})-Q(\hat{b}(\beta_0),\beta_0)\nonumber\\
&=&-(\tilde{Y}-\tilde{X}\beta_0)(\gamma_1\tilde{X}u+Z(Z^TZ)^{-1}v(\hat{\beta}))+(1/2)||\gamma_1\tilde{X}u+Z(Z^TZ)^{-1}v(\hat{\beta})||^2\nonumber\\
&&+(\tilde{Y}-\tilde{X}\beta_0)^TZ(Z^TZ)^{-1}v(\beta_0)-(1/2)||Z(Z^TZ)^{-1}v(\beta_0)||^2\nonumber\\
&&+n\lambda_1\sum_{j=1}^{p_1}w_{1j}||\hat{b}_j(\hat{\beta})||+n\lambda_1\sum_{j=p_1+1}^{s}w_{1j}\sqrt{K}|\hat{\beta}_j|+n\lambda_2\sum_{j=1}^{p_1}w_{2j}||\hat{b}_j(\hat{\beta})||_c\nonumber\\
&&-n\lambda_1\sum_{j=1}^{p_1}w_{1j}||\hat{b}_j(\beta_0)||-n\lambda_1\sum_{j=p_1+1}^{s}w_{1j}\sqrt{K}|\beta_{0j}|-n\lambda_2\sum_{j=1}^{p_1}w_{2j}||\hat{b}_j(\beta_0)||_c,\label{eqn:profile}
\end{eqnarray}
where for any random matrix $W$ with $n$ rows, we set $\tilde{W}=Q_ZW=W-P_ZW$ to be the projection of columns of $W$ onto the orthogonal complement of the column space of $Z$, where $P_Z=Z(Z^TZ)^{-1}Z^T$.

Using that $Z(Z^TZ)^{-1}Z'v$ is inside the column space of $Z$, while all variables with $\widetilde{\phantom{cc}} $ are orthogonal to it, the first four terms in (\ref{eqn:profile}) are simplified to 
\begin{eqnarray*}
-(\tilde{Y}-\tilde{X}\beta_0)^T(\gamma_1\tilde{X}u)+(1/2)||\gamma_1\tilde{X}u||^2+(1/2)||Z(Z^TZ)^{-1}v(\hat{\beta})||^2-(1/2)||Z(Z^TZ)^{-1}v(\beta_0)||^2.\\
\end{eqnarray*}
In Lemma \ref{lem:misc} (i)-(iii), we show that $||(\tilde{Y}-\tilde{X}\beta_0)^T(\tilde{X}u)||=O(\sqrt{ns})$, $||Z(Z^TZ)^{-1}v(\beta_0)||=O(\sqrt{nK}(\lambda_1||w'_1||+\lambda_2||w'_2||))$, and the last two lines in (\ref{eqn:profile}) involving the penalty terms is of order $O(n\sqrt{K}\lambda_1||w'_1||\gamma_1+nK(\lambda_1^2||w'_1||^2+\lambda_2^2||w'_2||^2)$. Since the eigenvalues of $\tilde{X}^T\tilde{X}/n$ are bounded away from zero by Lemma \ref{lem:misc} (iv) and condition (c6), $Q(\hat{b}(\hat{\beta}),\hat{\beta})-Q(\hat{b}(\beta_0),\beta_0)$ is bounded below by
\[nc\gamma_1^2+O(a_n)\gamma_1+O(b_n),\]
 for some $c>0$ and some positive sequences $a_n, b_n$, the exact expression of which we choose not to write down explicitly. Thus if $\gamma_1=C\max\{a_n/n,\sqrt{b_n/n}\}$ for $C>0$ sufficiently large, the above displayed expression will be positive. The expression $\max\{a_n/n,\sqrt{b_n/n}\}$ is exactly of order $\sqrt{s/n}+\sqrt{K}(\lambda_1||w'_1||+\lambda_2||w'_2||)$ as in the statement of the Theorem.  $\Box$ \\

\textit{Proof of Theorem \ref{thm:an}.} As in the proof of Theorem 1, $Z^{(1)}, X^{(2)}$ is simply written as $Z$ and $X$ here. By the KKT condition, in addition to that 
\begin{equation}\label{eqn:kkt1}
-Z_j^T(Y-Zb-X\beta)+n\lambda_1w_{1j}\partial||{b}_j||+n\lambda_2w_{2j}\partial||b_j||_c, j=1,\ldots,p_1,
\end{equation}
which has been used in the proof of Theorem \ref{thm:rates}, we also have that $(\hat{b},\hat{\beta})$ satisfies
\begin{equation}\label{eqn:kkt2}
-X_j^T(Y-Zb-X\beta)+n\lambda_1\sqrt{K}w_{1j}\partial|\beta_j|=0, j=p_1+1,\ldots,s.
\end{equation}

Since $Y=r'+X\beta+\epsilon$ where $r'=(r'_1,\ldots,r'_n)$ with $r_i'=\sum_{j=1}^{p_1}X_{ij}\beta_j(t)$, and denote by $b^0$ the vector containing the spline coefficients that achieve optimal approximation of $\beta_j(t), 1\le j\le p_1$, and set $a=r'-Zb^0$, (\ref{eqn:kkt2}) is rewritten as
\begin{equation*}
-X_j^T(\epsilon+a-Z(b-b^0)-X(\beta-\beta_0))+n\lambda_1w_{1j}\sqrt{K}\partial|\beta_j|=0, j=p_1+1,\ldots,s.
\end{equation*}
From (\ref{eqn:kkt1}), we get $Z(b-b^0)=Z(Z^TZ)^{-1}Z^T(\epsilon+a-X(\beta-\beta_0))+Z(Z^TZ)^{-1}v$ ($v=v(\beta)$ defined right after equation (\ref{eqn:defv})) and plug into the above displayed equation we get
\begin{eqnarray*}
-X_j^T(\epsilon+a-Z(Z^TZ)^{-1}[Z^T(\epsilon+a-X(\beta-\beta_0))+v]-X(\beta-\beta_0))&&\\
+n\lambda_1w_{1j}\sqrt{K}\partial|\beta_j|&=&0, j=p_1+1,\ldots,s,
\end{eqnarray*}
that is,
\begin{equation*}
-X_j^T(\widetilde{\epsilon+a}-\tilde{X}(\beta-\beta_0)-Z(Z^TZ)^{-1}v)+n\lambda_1w_{1j}\sqrt{K}\partial|\beta_j|=0, j=p_1+1,\ldots,s,
\end{equation*}
from which we get 
\begin{eqnarray}
&&\sqrt{n}\Sigma_n^{-1/2}A_n(\hat{\beta}-\beta_0)\nonumber\\
&=&\sqrt{n}\Sigma_n^{-1/2}A_n(\tilde{X}^T\tilde{X})^{-1}\tilde{X}^T(\epsilon+a)+\sqrt{n}\Sigma_n^{-1/2}A_n(\tilde{X}^T\tilde{X})^{-1}X^TZ(Z^TZ)^{-1}v\nonumber\\
&&+\sqrt{n}\Sigma_n^{-1/2}A_n(\tilde{X}^T\tilde{X})^{-1}\Lambda,\label{eqn:andecomp}
\end{eqnarray}
where $\Lambda$ is a $p_2-$dimensional vector with components given by $n\lambda_1w_{1j}\sqrt{K}\partial|\beta_j|, j=p_1+1,\ldots,s$. 
By Lemma \ref{lem:misc} (iv), we can replace $(\tilde{X}^T\tilde{X}/n)^{-1}$ by $\Xi^{-1}$ which only results in a multiplicative factor $1+o(1)$ and thus does not disturb the asymptotic distribution. 

It is easily shown
\[||\frac{1}{\sqrt{n}}\Sigma_n^{-1/2}A_n\Xi^{-1}||=O(\sqrt{s/n}).\]
Combining this with $||\tilde{X}^Ta||=O(\sqrt{ns/K^{2d}}+ns/K^{(d+d_g)})$ (combining bounds (\ref{eqn:2inlem})-(\ref{eqn:4inlem}) in Lemma \ref{lem:misc} (i) ), $||X^TZ(Z^TZ)^{-1}v||=O(n\sqrt{K}(\lambda_1||w'_1||+\lambda_2||w'_2||))$ (Lemma \ref{lem:misc} (ii) ) and $||\Lambda||=O(n\lambda_1\sqrt{K}||w'_1||)$, and conditions (c9)(c10), all terms in (\ref{eqn:andecomp}) are $o(1)$ except $\sqrt{n}\Sigma_n^{-1/2}A_n(\tilde{X}^T\tilde{X})^{-1}\tilde{X}^T\epsilon$, which can be shown to converge to $N(0,\sigma^2I)$ by Lindeberg-Feller central limit theorem using standard arguments. $\Box$.\\

\textit{Proof of Theorem \ref{thm:oracle}.}
Since $(\hat{b}^{(1)},\hat{\beta}^{(2)})$ solves the optimization problem (\ref{eqn:roe}), we have that 
\begin{eqnarray}
&&-Z_j^T(Y-Z^{(1)}\hat{b}^{(1)}-X^{(2)}\hat{\beta}^{(2)})+n\lambda_1w_{1j}\partial||\hat{b}^{(1)}_j||+n\lambda_2w_{2j}\partial||\hat{b}^{(1)}_j||_c=0, j=1,\ldots,p_1,\nonumber\\
				\label{eqn:kkt3}\\
&&-X_j^T(Y-Z^{(1)}\hat{b}^{(1)}-X^{(2)}\hat{\beta}^{(2)})+n\lambda_1w_{1j}\sqrt{K}\partial|\hat{\beta}^{(2)}_j|=0, j=p_1+1,\ldots,s.\label{eqn:kkt4}
\end{eqnarray}
We remind the readers that the equations above actually mean ``there exists some subdifferential that makes the left hand side zero" in case the subdifferential is not unique. 

In order to show that the $pK$-dimensional vector $\hat{b}=(\hat{b}^{(1)},\hat{b}^{(2)},\hat{b}^{(3)})$ with $\hat{b}^{(2)}_{jk}=\hat{\beta}^{(2)}_j,j=p_1+1,\ldots, s, k=1,\ldots, K$ and $\hat{b}^{(3)}_{jk}=0, s+1\le j\le p, 1\le k\le K$ solves (\ref{eqn:min2}), we only need to verify the corresponding KKT conditions, 
\begin{equation}\label{eqn:kktall}
-Z_j^T(Y-Z^{(1)}\hat{b}^{(1)}-Z^{(2)}\hat{b}^{(2)}-Z^{(3)}\hat{b}^{(3)})+n\lambda_1w_{1j}\partial||\hat{b}_j||+n\lambda_2w_{2j}\partial||\hat{b}_j||_c=0, j=1,\ldots,p.\\
\end{equation}

First, for $1\le j\le p_1$, (\ref{eqn:kktall}) trivially follows from (\ref{eqn:kkt3}), since $Z^{(2)}\hat{b}^{(2)}-Z^{(3)}\hat{b}^{(3)}=X^{(2)}\hat{\beta}^{(2)}$.

Next, for $p_1+1\le j\le s$, (\ref{eqn:kktall}) is implied by the following two results.
\begin{itemize}
\item[(a)] the $K-$dimensional vector $-Z_j^T(Y-Z^{(1)}\hat{b}^{(1)}-Z^{(2)}\hat{b}^{(2)}-Z^{(3)}\hat{b}^{(3)})+n\lambda_1w_{1j}\partial||\hat{b}_j||$ is orthogonal to $e:=(1,1,\ldots, 1)^T$.
\item[(b)] $||Z_j^T(Y-Z^{(1)}\hat{b}^{(1)}-Z^{(2)}\hat{b}^{(2)}-Z^{(3)}\hat{b}^{(3)})||+n\lambda_1w_{1j}\le n\lambda_2w_{2j}$.
\end{itemize}
In fact, (a) implies that $-Z_j^T(Y-Z^{(1)}\hat{b}^{(1)}-Z^{(2)}\hat{b}^{(2)}-Z^{(3)}\hat{b}^{(3)})+n\lambda_1w_{1j}\partial||\hat{b}_j||=Q_La$ ($Q_L$ is the matrix of projection onto the orthogonal complement of $e$ as defined in Section 2) for some $a$, and (b) implies that $||Q_La||\le n\lambda_2w_{2j}$ and thus we can find a version of $a$ with $||a||\le n\lambda_2w_{2j}$. If we choose the subdifferential $\partial ||\hat{b}_j||_c$ to be $-Q_La/(n\lambda_2w_{2j})$ (note that this is indeed a subdifferential since $||\hat{b}_j||_c=0$ when $p_1+1\le j\le s$) then equation (\ref{eqn:kktall}) is verified.

For verifying (a), we can set $\partial ||\hat{b}_j||=(s_j,\ldots,s_j)/\sqrt{K}$ where $s_j=\partial |\beta^{(2)}_j|$ in (\ref{eqn:kkt4}) (it can be verified that $(s_j,\ldots,s_j)/\sqrt{K}$ is indeed a subdifferential). With this choice of $\partial ||\hat{b}_j||$ , it can be easily checked that
$e^T\{-Z_j^T(Y-Z^{(1)}\hat{b}^{(1)}-Z^{(2)}\hat{b}^{(2)}-Z^{(3)}\hat{b}^{(3)})+n\lambda_1w_{1j}\partial||\hat{b}_j||\}$ is exactly equal to the left hand side of (\ref{eqn:kkt4}) and thus equal to zero, which immediately implies (a).

For verifying (b), we have $||Z_j^T(Y-Z^{(1)}\hat{b}^{(1)}-Z^{(2)}\hat{b}^{(2)}-Z^{(3)}\hat{b}^{(3)})||\le||Z_j^T\epsilon||+||Z_j^T(Z^{(1)}(\hat{b}^{(1)}-b^0)+X^{(2)}(\hat{\beta}^{(2)}-\beta_0^{(2)}))||+||Z_j^T(r'-Z^{(1)}b^0)||$
($r'$, $b^0$ defined in the proof of Theorem \ref{thm:an}). Using exactly the same arguments as in Theorem 1 of \cite{huang10}, we have $\max_j||Z_j^T\epsilon||=O(\sqrt{(n/K)\log(pK)})$. Besides, it is easy to see (using Theorem \ref{thm:rates}) that $||Z_j^T(Z^{(1)}(\hat{b}^{(1)}-b^0)+X^{(2)}(\hat{\beta}^{(2)}-\beta_0^{(2)}))||+||Z_j^T(r'-Z^{(1)}b^0)||=O\left(\sqrt{(n/K)(Ks+ns/K^{2d}+nK(\lambda_1^2||w'_1||^2+\lambda_2^2||w'_2||^2))}\right)$ and (b) is verified by condition (c11) (condition (c11) also implies $\lambda_1||w'_{1}||=o(\lambda_2w_{2j}), p_1+1\le j\le s$). 

Finally, for $s+1\le j\le p$ in (\ref{eqn:kktall}), we only need to verify that $||Z_j^T(Y-Z^{(1)}\hat{b}^{(1)}-Z^{(2)}\hat{b}^{(2)}-Z^{(3)}\hat{b}^{(3)})||\le n\lambda_1w_{1j}, s+1\le j\le p$ which follows exactly the same arguments
as in verifying (b) above and the details are omitted. $\Box$\\

\textit{Proof of Theorem \ref{thm:bic}.}
For any given pair of regularization parameters $\lambda=(\lambda_1,\lambda_2)$, we denote by $\hat{b}_\lambda$ the minimizer of (\ref{eqn:min2}), and by $\hat{b}$ the minimizer when the optimal sequence of regularization parameters is chosen such that $\hat{b}$ results in a consistent model selection. We separately consider several different cases below. For each case, we implicitly assume that all previous cases do not happen since they have already been dealt with. 

\textit{Case 1. Some truly varying coeffients are estimated as constant or zero coefficients in $\hat{b}_\lambda$.} Similar to the calculations performed in the proof of Theorem \ref{thm:rates}, we have
\begin{eqnarray*}
&&\frac{1}{2n}||Y-Z\hat{b}_\lambda||^2-\frac{1}{2n}||Y-Z\hat{b}||^2\\
&\ge&-\frac{1}{n}||P_Z(Y-Z\hat{b})||^2+\frac{1}{4n}||Z(\hat{b}-\hat{b}_\lambda)||^2.
\end{eqnarray*}
Since there is some $j$ for which $\hat{b}_j$ represents a truly varying coefficient with convergence rate given by Theorem \ref{thm:rates}, while $\hat{b}_{\lambda j}$ has all $K$ components equal to each other representing a constant coefficient, it is easy to show that $||Z(\hat{b}-\hat{b}_\lambda)||^2/n\ge ||Z_j(\hat{b}_j-\hat{b}_{\lambda j})||^2/n$ is bounded away from zero by condition (c13). Besides, $||P_Z(Y-Z\hat{b})||/n=o(1)$ (using the same arguments as in Lemma \ref{lem:eta} as well as the proof of convergence rate in Theorem \ref{thm:rates}) and the penalty terms in BIC are all of order $o(1)$, thus the BIC when $\lambda$ is used is bigger than the BIC when the optimal regularization sequence is used (following the same arguments as in the proof of Theorem ? in \cite{?}). 
 
\textit{Case 2. Some  nonzero constant coefficients are estimated as zeros in $\hat{b}_\lambda$.} This also represents an underfitted model and is dealt with similarly as in Case 1.

\textit{Case 3. Some zero or constant coefficients are estimated as truly varying in $\hat{b}_\lambda$.} Let $\hat{b}^*$ be the minimizer of the least square $||Y-Zb||^2$ under the additional constraint that the model identified by $\hat{b}_\lambda$ is used when minimizing the least square. We have that
\begin{eqnarray}
&&\frac{1}{2n}||Y-Z\hat{b}_\lambda||^2-\frac{1}{2n}||Y-Z\hat{b}||^2\nonumber\\
&\ge&\frac{1}{2n}||Y-Z\hat{b}^*||^2-\frac{1}{2n}||Y-Z\hat{b}||^2\nonumber\\
&=&\frac{1}{n}(Y-Z\hat{b})^TZ(\hat{b}-\hat{b}^*)+\frac{1}{2n}||Z(\hat{b}-\hat{b}^*)||^2\nonumber\\
&\ge&\frac{1}{n}(Y-Z\hat{b})^TZ(\hat{b}-\hat{b}^*).\label{eqn:overfit}
\end{eqnarray}
By the definition of $\hat{b}^*$ and the fact that we only search over models with size $O(s)$, the convergence rate of $\hat{b}^*$ can be obtained using similar arguments as Theorem \ref{thm:rates} but without the terms involving $\lambda_1$ and $\lambda_2$ appearing. Arguments similar to those used in showing result (b) in the proof of Theorem \ref{thm:oracle} can be used to show that the (\ref{eqn:overfit}) is bounded below by a negative term whose absolute value is of order
\[\frac{1}{n}\sqrt{(ns\log(pK)+\frac{n^2s}{K^{2d+1}})\cdot(\frac{K^2s}{n}+\frac{s}{K^{2d-1}})}=O(\frac{Ks\sqrt{\log(pK)}}{n}),\]
which is of order smaller than the BIC penalty term $\log (n/K)/(n/K)C_n$ when $C_n=\Omega(\sqrt{\log(pK)})$ (note we assume $s=O(1)$). That BIC cannot select such $\lambda$ can now be derived by standard arguments.

\textit{Case 4. Some zero coefficients are estimated as nonzero constants. } This case is similar to the previous one and the details are omitted. $\Box$\\

\textit{Proof of Theorem \ref{thm:init}.} We only sketch the proof here. First using the general results in \cite{weihuang07}, which deal with linear models with group Lasso penalty, we can show that at most $O(s)$ covariates are selected if $\lambda_0>\sqrt{s\log(pK)/n}$. The only difference of our case from that of \cite{weihuang07} is the necessity of an approximation of coefficient functions by spline expansions. However, this problem can be solved by following exactly the same lines in the proof of Theorem 1 in \cite{huang10}, using the bound for $||r-Zb^0||$ in Lemma \ref{lem:eta}. The rest of the proof on convergence rate follows the same strategy as in Theorem \ref{thm:rates}.

\begin{lem}\label{lem:eta}
Following notations defined in the proof of Theorem \ref{thm:rates}, $||\eta||^2=||P_Z(Y-Zb^0)||^2$ is of order $O(Ks+ns/K^{2d})$.
\end{lem}
\textit{Proof.} Denote $r_i=\sum_{j=1}^sX_{ij}\beta_j(t_i)$ and $r=(r_1,\ldots,r_n)^T$. We have $Y-Zb^0=\epsilon+(r-Zb^0)$ and $||\eta||^2\le 2||P_Z\epsilon||^2+2||r-Zb^0||^2$. By the approximation property of splines, $||r-Zb^0||^2=O_p(ns/K^{2d})$. Also, $E||P_Z\epsilon||^2=E(\epsilon^TP_Z\epsilon)=\sigma^2 tr(P_Z)=O(sK)$ and the lemma is proved by an application of Markov inequality. $\Box$\\

We collect several miscellaneous results on bounding some terms used in the proof of Theorem \ref{thm:rates} and Theorem \ref{thm:an} in the following Lemma.
\begin{lem}\label{lem:misc} Following the notations used in Theorem \ref{thm:rates} and Theorem \ref{thm:an}, we have
\begin{itemize}
\item[(i)] $||(\tilde{Y}-\tilde{X}\beta_0)^T\tilde{X}||=O(\sqrt{ns})$.
\item[(ii)] $||Z(Z^TZ)^{-1}v(\beta_0)||=O(\sqrt{nK}(\lambda_1||w'_1||+\lambda_2||w'_2||))$.
\item[(iii)] The last two lines in (\ref{eqn:profile}) is of order $O(n\sqrt{K}\lambda_1||w'_1||\gamma_1+nK(\lambda_1^2||w'_1||^2+\lambda_2^2||w'_2||^2))$.
\item[(iv)] $||\tilde{X}^T\tilde{X}/n-\Xi||=o(1)$ where $||B||$ for a matrix $B$ denotes its Frobenius norm.
\end{itemize}
\end{lem}
\textit{Proof.} \\

(i) We first write down the decomposition  
\[X=\Theta-G+G+U\] 
(note we follow the notation in Theorem 1 and 2 and write $X^{(2)}$ simply as $X$).
The above uppercase letters represent $n\times p_2$ matrices, and correspond to the decomposition in (\ref{eqn:decomp}) evaluated at $n$ observations.
After projection, we have
\[\tilde{X}=\tilde{\Theta}-\tilde{G}+\tilde{G}+\tilde{U}.\] 
Together with the decomposition 
\[\tilde{Y}-\tilde{X}\beta_0=\tilde{\epsilon}+\widetilde{(r'-Zb^0)},\]
(same as in the proof of Theorem \ref{thm:an}, $r'=(r'_1,\ldots,r'_n)^T$ with $r_i'=\sum_{j=1}^{p_1}X_{ij}\beta_j(t)$, $b^0$ contains the spline coefficients that achieve optimal approximation of $\beta_{0j}(t), 1\le j\le p_1$), the bound for $||(\tilde{Y}-\tilde{X}\beta_0)^T\tilde{X}||$ is obtained from the following estimates.
\begin{eqnarray}
||\epsilon^TQ_ZX||&=&O(\sqrt{ns}), \label{eqn:1inlem}\\
||(r'-Zb^0)^TQ_Z(\Theta-G)||&=&\sqrt{\frac{ns}{K^{2d}}}, \label{eqn:2inlem}\\
||(r'-Zb^0)^TQ_ZU||&=&\sqrt{\frac{ns}{K^{2d}}},\label{eqn:3inlem}\\
||(r'-Zb^0)^TQ_ZG||&=&\sqrt{\frac{ns}{K^{2d}}}\sqrt{\frac{ns}{K^{2d_g}}}=O(\sqrt{ns}),\label{eqn:4inlem}
\end{eqnarray}
where (\ref{eqn:1inlem}) is obvious from condition (c1), (\ref{eqn:2inlem}) is based on that entries of $\Theta-G$ have mean zero and are orthogonal to $\mathcal{G}$ while entries of $(r'-Zb^0)^T$ and $Z$ are inside $\mathcal{G}$ and thus we can calculate the bound by considering its variance, (\ref{eqn:3inlem}) is obtained similarly, and finally $(\ref{eqn:4inlem})$ is obtained from $||Q_ZG||\le||G||=O(\sqrt{ns/K^{2d_g}})$ and conditions (c8).\\

(ii) Obviously $||Z(Z^TZ)^{-1}v(\beta)||^2=O(K/n)||v(\beta)||^2$. Using the fact that $\partial||\hat{b}_j||$ and $\partial||\hat{b}_j||_c$ has $l_2$ norm bounded by 1, it easily follows from the definition of $v(\beta)$ (below equation (\ref{eqn:defv})) that $||v(\beta_0)||^2=O(n^2(\lambda_1^2||w'_1||^2+\lambda_2^2||w'_2||^2))$.\\

(iii) We have
\begin{eqnarray*}
&&n\lambda_1\sum_{j=1}^{p_1}w_{1j}||\hat{b}_j(\hat{\beta})-\hat{b}_j(\beta_0)||\\
&\le&n\lambda_1||w'_1||\cdot ||\hat{b}(\hat{\beta})-\hat{b}(\beta_0)||\\
&\le&n\lambda_1||w'_1||\cdot(||(Z^TZ)^{-1}Z^T(\hat{\beta}-\beta_0)||+||(Z^TZ)^{-1}(v(\hat{\beta})-v(\beta_0))||)\\
&=&n\lambda_1||w'_1||(\gamma_1\sqrt{K/n}+K(\lambda_1||w'_1||+\lambda_2||w'_2||)\\
&=&O(\sqrt{nK}\lambda_1||w'_1||\gamma_1+nK(\lambda_1^2||w'_1||^2+\lambda_2^2||w'_2||^2)),
\end{eqnarray*}
where in the 2nd line above we used Cauchy-Schwartz inequality, in the 3rd line we used (\ref{eqn:defv}), in the 4th line we used part (ii) of this Lemma. We can bound  $n\lambda_2\sum_{j=1}^{p_1}w_{2j}||\hat{b}_j(\hat{\beta})-\hat{b}_j(\beta_0)||_c$ in a similar way.

Finally, 
\begin{eqnarray*}
&&n\lambda_1\sqrt{K}\sum_j\{w_{1j}(|\hat{\beta}_j|-|\beta_{0j}|)\}\\
&\le&n\lambda_1\sqrt{K}\sum_j\{w_{1j}|\hat{\beta}_j-\beta_{0j}|\}\\
&\le&n\lambda_1\sqrt{K}||w'_1||\gamma_1,
\end{eqnarray*}
using Cauchy-Schwartz inequality in the last line above.

(iv) Using the decomposition $\tilde{X}=\Gamma-P_Z\Gamma+\tilde{G}+U-P_ZU$ where $\Gamma=\Theta-G$, we have that 
\begin{equation}\label{eqn:iv1}
||\frac{(\Gamma+U)^T(\Gamma+U)}{n}-\Xi||=O(\frac{s}{\sqrt{n}})=o(1),
\end{equation}
since each entry of $(\Gamma+U)^T(\Gamma+U)/n-\Xi$ has mean zero and the above can be proved by calculating the variance of each entry (this is just a standard way of proving the weak law of large numbers). 

We also have the following bounds.
\begin{equation}
||\frac{\Gamma^T P_Z \Gamma}{n}||=O(\frac{s}{n}tr(P_Z))=O(\frac{s^2K}{n}),
\end{equation}
by that each entry of $\Gamma$ is orthogonal to $\mathcal{G}$ and entries of $Z$ are in $\mathcal{G}$. 
\begin{equation}
||\frac{U^T P_Z U}{n}||=O(\frac{s}{n}tr(P_Z))=O(\frac{s^2K}{n}),
\end{equation}
by a similar reason as before.

\begin{equation}\label{eqn:iv4}
||\frac{G^T Q_Z G}{n}||=O(\frac{s^2}{K^{2d_g}})
\end{equation}
by condition (c7).

Other terms in $||\tilde{X}^T\tilde{X}/n-\Xi||$ can be bounded by Cauchy-Schwartz inequality utilizing (\ref{eqn:iv1})-(\ref{eqn:iv4}), resulting in some additional $o(1)$ terms, and part (iv) of the Lemma is proved.

\bibliographystyle{jasa}
\bibliography{papers.txt,books.txt}

\end{document}